\documentclass[12pt]{article}
\usepackage{graphicx}

\newsavebox{\sboxpubnumber}
\newsavebox{\sboxpubdate}
\newcommand{\pubdate}[1]{\begin{lrbox}{\sboxpubdate}{#1}\end{lrbox}}
\newcommand{\pubnumber}[1]{\begin{lrbox}{\sboxpubnumber}{\begin{tabular}{l} #1 \\
\usebox{\sboxpubdate}
\end{tabular}}
                           \end{lrbox}
                           \pubblock}
\newcommand{\Title}[1]{\begin{center} {\Large #1 } \end{center}}
\newcommand{\Author}[1]{\begin{center}{ \sc #1} \end{center}}
\newcommand{\Address}[1]{\begin{center}{ \it #1} \end{center}}

\newcommand{\pubblock}{\rightline{
\usebox{\sboxpubnumber}}}

\newcommand{\bea}{\begin{eqnarray}}
\newcommand{\eea}{\end{eqnarray}}

\newenvironment{Abstract}{\begin{quotation}  }{\end{quotation}}
\newenvironment{Presented}{\begin{quotation} \begin{center}
             PRESENTED AT\end{center}\bigskip
      \begin{center}\begin{large}}{\end{large}\end{center}
      \end{quotation}}
\newcommand{\Acknowledgements}{\bigskip  \bigskip \begin{center} \begin{large}
             \bf ACKNOWLEDGEMENTS \end{large}\end{center}}

\begin{document}

\begin{titlepage}
\pubdate{\today}                    
\pubnumber{XXX-XXXXX \\ YYY-YYYYYY} 

\vfill
\Title{
CMBR constraints on $R^2$ gravity}
\vfill
\Author{Hyerim Noh
                          }
\Address{Korea Astronomy Observatory, Korea, 
 \\
         and Institute of Astronomy, Univ. of Cambridge, UK}
\vfill
\vfill
\vfill
\begin{Abstract}
Considering the inflation model based on a $f(R)$ gravity theory, we
obtain several important constraints
from the large angular scale CMBR observations.
First, the ordinary slow-roll assumption during the inflation  together
with Harrison-Zel'dovich spectral conditions chooses $R^2$ gravity as a unique
candidate.
Second, the $R^2$ gravity leads to specific near scale-invariant 
Harrison-Zel'dovich spectra both for the scalar and the tensor perturbations.
Third, using the COBE-DMR data we derive the strong constraints
on the coupling constant and the energy scale during the inflation.
Also, our result shows the gravitational wave contribution to the
CMBR anisotropy is negligible.
So, the future observation can provide the strong constraints on the
the inflation model based on $R^2$ gravity.
This is a summary of a talk presented in COSMO-01, and the
more complete published version can be found in astro-ph/0102423.

\end{Abstract}
\vfill
\begin{Presented}
    COSMO-01 \\
    Rovaniemi, Finland, \\
    August 29 -- September 4, 2001
\end{Presented}
\vfill
\end{titlepage}
\def\thefootnote{\fnsymbol{footnote}}
\setcounter{footnote}{0}

\section{Introduction}
The inflation stage in the early universe provides a natural mechanism
which generates the origin of the large scale structures and the
gravitational wave background.
Thus, naturally, the observed large-scale structures and the 
gravitational wave background can constrain the model parameters 
of the proposed inflation models.
In this context, the cosmic microwave background radiation (CMBR) 
temperature and polarization anisotropies in the
large angular scale provide important constraints
on the inflation models which are usually based on the
scalar field or the generalized gravity theories.
There have been many attempts to constrain the model parameters of
simple models based on a single field potential in Einstein gravity.
Recenlty, there have been growing interests on the roles of the
generalized gravity theories other than the Einstein gravity
in the early universe.
These generalized gravity theories naturally arise either from
attempts to quantize the gravity or as the low-energy
limits of the unified theories including gravity.
Especially, the $R^2$ terms appear in many theories like
Kaluza-Klein models, string/M-theory programs, etc.
The $R^2$ gravity has a natural reheating mechanism, so
provides a self-contained inflation model without introducing
a field or a phase transition \cite{Starobinsky:1980}.

We use the CMBR temperature anisotropy in the large angular scale, 
and derive constraints on the
inflationary models based on the $f(R)$ gravity theory.
We consider general $f(R)$ term in the Lagrangian and show
that the Harrison-Zel'dovich spectrum \cite{Harrison-Zeldovich}
with the slow roll assumption during the
inflation  chooses the $R^2$ gravity as a unique candidate.
Our derived constraints include the coupling constant of
$R^2$ term and the energy scale during the inflation.
We also show that the gravitational wave contribution to the
CMBR anisotropy is suppressed.

\section{Classical evolution}

We consider a Lagrangian \cite{HN:1996,HN:1998}
\bea
   & & L =  {1 \over 2} f (\phi, R)
       - {1\over 2} \omega (\phi) \phi^{;a} \phi_{,a} - V(\phi),
   \label{Lagrangian}
\eea
where $f(\phi, R)$ is a general algebraic function of the scalar field $\phi$
and the scalar curvature $R$; $\omega (\phi)$ and $V (\phi)$ are
general algebraic functions of $\phi$.
We introduce $F \equiv  \partial f/\partial R$ .

We consider a spatially homogeneous and isotropic Friedmann world model
together with the most general scalar- and tensor-type spacetime
dependent perturbations
\cite{Bardeen-1988}
\bea
   d s^2
   &=& - a^2 \left( 1 + 2 \alpha \right) d \eta^2
       - 2a^2 \beta_{,\alpha} d \eta d x^\alpha
   \nonumber \\
   & & + a^2 \left( \delta_{\alpha\beta}
       + 2 \varphi \delta_{\alpha\beta} + 2 \gamma_{,\alpha\beta}
       + 2 C_{\alpha\beta} \right) d x^\alpha d x^\beta,
   \label{metric-general}
\eea
where
the transverse-tracefree $C_{\alpha\beta}$ indicates the gravitational wave.
We also decompose $\phi = \bar \phi + \delta \phi$,
$F = \bar F + \delta F$, etc.

The perturbed action in a unified form is given by \cite{HN:1998}
\bea
   & & \delta^2 S = {1 \over 2} \int a^3 Q \left( \dot \Phi^2
       - {1 \over a^2} \Phi^{|\gamma} \Phi_{,\gamma}
       \right) dt d^3 x.
   \label{perturbed-action}
\eea

This applies for  the second order gravity such as either
$f=f(\phi )R$ in the presence of $\phi$ or $f=f(R)$ without a field.
{}For the scalar- and tensor-type perturbations \cite{HN:1996,HN:1998}
\bea
   & & \Phi = \varphi_{\delta \phi}, \quad
       Q = { \omega \dot \phi^2 + {3 \dot F^2 \over 2 F}
       \over \left( H + {\dot F \over 2 F} \right)^2 },
   \label{case-field} \\
   & & \Phi = C^\alpha_\beta, \quad \; Q = F ,
   \label{case-GW}
\eea
where
$H\equiv \dot a/a$, $\varphi_{\delta\phi } \equiv \varphi -H\delta\phi
/\dot\phi$ is the gauge invariant variable.

The equation of motion and the large scale solution are given by
\cite{HN:1996,HN:1998}
\bea
   & & {1 \over a^3 Q} (a^3 Q \dot \Phi)^\cdot
       - {\Delta \over a^2} \Phi = 0,
   \label{Phi-eq} \\
   & & \Phi = C ({\bf x}) - D ({\bf x}) \int_0^t {dt \over a^3 Q}.
   \label{Phi-LS-sol}
\eea
Ignoring the transient solution we have a temporally conserved behavior
\bea
   & & \Phi ({\bf x}, t) = C ({\bf x}).
   \label{conservation}
\eea

Using $z \equiv a \sqrt{Q}$ and $v \equiv z \Phi$ eq. (\ref{Phi-eq}) becomes
\bea
   & & v^{\prime\prime} + \left(  k^2 - z^{\prime\prime} / z \right) v
       = 0,
   \label{v-eq}
\eea
where a prime denotes the time derivative with respect to $\eta$.

\section{Quantum generation}

We introduce the slow-roll parameters \cite{HN:1996}
\bea
   & & \epsilon_1 \equiv {\dot H \over H^2}, \quad
       \epsilon_2 \equiv {\ddot \phi \over H \dot \phi}, \quad
       \epsilon_3 \equiv {1 \over 2} {\dot F \over H F}, \quad
       \epsilon_4 \equiv {1 \over 2} {\dot E \over H E},
\eea
where
  $E \equiv F \left( \omega + {3 \dot F^2 \over 2 \dot \phi^2 F} \right)$.

{\it Assuming} $\dot \epsilon_i = 0$ we have
$z^{\prime\prime}/z = n/\eta^2$ where for the scalar- ($n=n_s$) and
tensor ($n=n_t$) perturbations \cite{HN:1996}
\bea
   & & n_s = { ( 1 - \epsilon_1 + \epsilon_2 - \epsilon_3 + \epsilon_4 )
       ( 2 + \epsilon_2 - \epsilon_3 + \epsilon_4 )
       \over ( 1 + \epsilon_1 )^2 },
   \nonumber \\
   & & n_t = { ( 1 + \epsilon_3 ) ( 2 + \epsilon_1 + \epsilon_3 )
       \over (1 + \epsilon_1)^2 }.
   \label{n}
\eea
{}For constant $n$, above equation becomes a Bessel's equation.

In the large-scale limit, the general power spectra
based on vacuum expectation values lead to \cite{HN:1998} 
\bea
   & & {\cal P}_{\hat \varphi_{\delta \phi}}^{1/2}
       = {1\over \sqrt{Q}}{H\over 2\pi}
         {1\over aH|\eta |}
         {\Gamma(\nu_s) \over \Gamma(3/2)}
       \left( { k |\eta| \over 2 } \right)^{3/2-\nu_s},
   \label{P-s-NM-q} \\
   & & {\cal P}_{\hat C_{\alpha\beta}}^{1/2}
       = \sqrt{{2 \over Q}}{H\over 2\pi} {1 \over aH|\eta |}
       {\Gamma(\nu_t) \over \Gamma(3/2)}
       \left( { k |\eta| \over 2 } \right)^{3/2-\nu_t},
   \nonumber \\
   \label{P-GW-NM-q}
\eea
where
$\nu_s \equiv \sqrt{n_s +1/4}$ and
$\nu_t \equiv \sqrt{n_t +1/4}$,
and we neglected the dependences on the vacuum choice.

\section{Observations and the spectral constraints}

In \cite{HN:1996} it is shown that
$\varphi_{\delta \phi}$ and $C_{\alpha\beta}$ are generally
conserved independently of changing gravity  theory,
changing potential, and changing equation of state, as long as the scale
remains in the large-scale; this applies to the  case of the
observationally relevant scales before the second horizon crossing
in the matter dominated era.
Using these conserved properties, we can identify the power spectra
based on the vacuum expectation value during the inflation era
[${\cal P}_{\hat \varphi_{\delta \phi}}$ and
${\cal P}_{\hat C_{\alpha\beta}}$]
with the classical power spectra based on the spatial average
[${\cal P}_{\varphi_{\delta \phi}}$ and ${\cal P}_{C_{\alpha\beta}}$].
Then, we have the same results in
eqs. (\ref{P-s-NM-q},\ref{P-GW-NM-q}) for
${\cal P}_{\varphi_{\delta \phi}}$ and ${\cal P}_{C_{\alpha\beta}}$.
Thus, eqs. (\ref{P-s-NM-q},\ref{P-GW-NM-q})
remain valid even in the matter dominated era.
The spectral indices of the scalar- and tensor-type structures are
given as $n_S -1 = 3 - \sqrt{4n_s +1}$ and $n_T = 3 - \sqrt{4n_t +1}$.
{}For the scale independent spectra the quadrupole anisotropy is 
\cite{HN:1998-PRL}
\bea
   \langle a_2^2 \rangle
   &=& \langle a_2^2 \rangle_S + \langle a_2^2 \rangle_T
   \nonumber \\
   &=& {\pi \over 75} {\cal P}_{\varphi_{\delta \phi}}
       + 7.74 {1 \over 5} {3 \over 32} {\cal P}_{C_{\alpha\beta}}.
   \label{a_l}
\eea
Then the ratio between the two perturbations becomes
\bea
   & & r_2 \equiv { \langle a_2^2 \rangle_T \over \langle a_2^2 \rangle_S }
       = 3.46 { {\cal P}_{C_{\alpha\beta}} \over
       {\cal P}_{\varphi_{\delta \phi}} }.
   \label{r_2-original}
\eea
The four-year {\it COBE}-DMR data provide \cite{Gorski:1998}
\bea
   & & \langle a_2^2 \rangle \simeq 1.1 \times 10^{-10}.
   \label{a_2-value}
\eea

Now, we consider a special case with 
\bea
   & & L = {1 \over 2} f (R).
   \label{Lagrangian-f}
\eea
The background equations are \cite{HN:1996}
\bea
   & & H^2 = {1 \over 3F} \left( {RF - f \over 2} - 3 H \dot F \right),
   \label{BG1} \\
   & & \dot H = - {1\over 2 F} \left( \ddot F - H \dot F \right),
   \label{BG2}
\eea
with $R = 6 ( 2 H^2 + \dot H )$.
Equation (\ref{BG2}) gives
\bea
   & & \epsilon_1 = \epsilon_3 ( 1 - \epsilon_4).
   \label{epsilon_34}
\eea
As the slow-roll conditions,
we assume
\bea
   & & |\epsilon_1| \ll 1.
   \label{slow-roll}
\eea
We have $\epsilon_2 = 0$.
The condition
$\dot \epsilon_i = 0$ is reasonable in the sense that
the large-scale structures are generated from
a short duration (about 60 $e$-folds) of the inflation  and the time
variation of $\epsilon_i$ during that period is negligible.
{}For constant $\epsilon_i$ we have
$f \propto R^{(2 + \epsilon_1)/(1 + \epsilon_1 - 2 \epsilon_3)}$.
{}From eqs. (\ref{epsilon_34},\ref{slow-roll}) we have
$\epsilon_3 \simeq 0$ or $\epsilon_4 \simeq 1$.
In the case of $\epsilon_4 \simeq 1$, 
the spectral constraint $n_S - 1 \simeq 0$
, thus $n_s \simeq 2$ in
eq. (\ref{n}) leads to $\epsilon_3 \simeq 1$ or $4$. Both are excluded
if we use $n_T \simeq 0$ thus $n_t \simeq 2$ in eq. (\ref{n}).
So, we have $\epsilon_3 \ll 1$ and
\bea
   & & f \propto R^{2 - \epsilon_1 + 4 \epsilon_3}.
   \label{fR-slow-roll}
\eea
Using $\epsilon_3 \simeq 0$ in eq. (\ref{n}) with $n_s \simeq 2$
gives $\epsilon_4 \simeq 0$ or $-3$; correspondingly we have
$\epsilon_1 \simeq \epsilon_3$ or $4 \epsilon_3$.
{}For $\epsilon_4 \simeq -3$ we have exactly
$f = R^2$ to the linear order in $\epsilon_i$.
If $\epsilon_4 \simeq 0$, then 
$\epsilon_3 = \epsilon_1$, thus
\bea
   & & f \propto R^{2 + 3 \epsilon_1}.
   \label{pure-R^2}
\eea

Our result shows that near scale-invariant Harrison-Zel'dovich spectra for
{\it both} scalar- and tensor-type structures
generated from inflation based on $f(R)$ gravity
choose $f \propto R^2$ gravity.
However, the current observations of CMBR provide a constraint on only the
sum of the two types of structures.
Usually, the observations provide constraint on the
scalar-type perturbation, thus its amplitude and the spectral index.
In such a case, our adoption of the Harrison-Zel'dovich type spectral 
index for the
tensor-type structure as well to sort out the $R^2$ may be regarded as
having a loop-hole in the argument.
However, since we have used the $n_T$ constraint to exclude
the cases of $\epsilon_4 \simeq 1$, thus $\epsilon_3 \simeq 1$ or $4$
just above eq. (\ref{fR-slow-roll})
we have $f \propto R^{-2}$ or $R^{-2/7}$,
respectively.
The corresponding values of the
spectral index for the tensor-type perturbation are $n_T \simeq -2$ or $-8$,
respectively, which are far from the expected values.

{}For $|\epsilon_i| \ll 1$ the spectral indices
of the scalar- and the tensort-type perturbations become
\bea
   & & n_S - 1 \simeq 2 ( 3 \epsilon_1 - \epsilon_4 ), \quad
       n_T \simeq 0,
   \label{n-result-f}
\eea
to $\epsilon_i$-order.
\section{COBE-DMR constraints}
                                             \label{sec:R^2}

We consider the following gravity during inflation
\bea
   & & f = {m_{pl}^2 \over 8 \pi} \left( R + {R^2 \over 6 M^2} \right).
   \label{Lagrangian-R^2}
\eea
$R^2$ inflation model has been investigated extensively
\cite{Starobinsky:1980,R-square}.
During inflation we {\it assume} the second term dominates over the
Einstein action which {\it requires} $H^2_{\rm infl} \gg M^2$;
this is a slow-roll regime and we have $|\epsilon_i| \ll 1$.
{}For the model in eq. (\ref{pure-R^2}), the pure $R^2$ gravity implies
$\epsilon_1 = 0$ and exponential inflation.
However, in our case of eq. (\ref{Lagrangian-R^2})
we do have $\epsilon_1$.
In the slow-roll regime we have
$\epsilon_3 = \epsilon_1$, thus eq. (\ref{n-result-f}) remains valid.
Assuming $H^2 \gg M^2$ we derive the power spectra
\bea
   & & {\cal P}^{1/2}_{\varphi_{\delta F}}
       = {1 \over 2 \sqrt{3 \pi}} {1 \over |\epsilon_1|} {M \over m_{pl}},
       \quad
       {\cal P}^{1/2}_{C_{\alpha\beta}}
       = {1 \over \sqrt{\pi}} {M \over m_{pl}},
   \label{P-spectra}
\eea
where we ignored $[ 1 + {\cal O} (\epsilon_i)]$ factors.

Then, we have
\bea
   & & r_2 = 41.6 \epsilon_1^2.
   \label{r}
\eea
Comparing with the minimally coupled scalar field case where
$r_2 = - 6.93 n_T$ called consistency relation, in $R^2$ gravity
$r_2$ in eq. (\ref{r}) is quadratic order in $\epsilon_1$, and
$n_T$ in eq. (\ref{n-result-f}) vanishes to the linear order in $\epsilon_i$.
Therefore, in order to check
whether the consistency relation remains in the $R^2$ gravity
we need to consider the second-order effects of the slow-roll parameters.

{} Eqs. (\ref{a_l},\ref{P-spectra},\ref{a_2-value}) give
\bea
   & & {M \over m_{pl}} = 3.1 \times 10^{-4} |\epsilon_1|.
   \label{M-constraint}
\eea
Many works have shown the similar constraint \cite{R-square}.
While the  previous works are based on an asymptotic solution
of the background $R^2$-gravity inflation model,
ours are based on exact solutions available when $\dot \epsilon_i = 0$.
We can handle the perturbations
analytically and derive the solution without using the
conformal transformation.

{}From eqs. (\ref{BG1},\ref{BG2}) we derive
$ \dot H^2 - 6 H^2 \dot H - H^2 M^2 - 2 H \ddot H = 0$.
In the limit $H^2 \gg M^2$, and with the neglible $\ddot H$ term,
we have a solution during the inflation as $H = H_i - M^2 (t - t_i)/6$.
Assuming that the inflation ends near $(t_e - t_i) \sim 6 H_i/M^2$,
we can estimate the number of $e$-folds from $t_k$ (when the relevant scale
$k$ exits the Hubble horizon and reaches the large-scale limit)
till the end of inflation as
\bea
   & & N_k \equiv \int_{t_k}^{t_e} H dt = - {H^2 \over 2 \dot H} (t_k)
       = - {1 \over 2 \epsilon_1 (t_k)}.
   \label{N_k}
\eea
Thus $H(t_k)/M = \sqrt{N_k / 3}$.
Since the large-angular CMBR scales exit the horizon about
$50 \sim 60$ $e$-folds before the end of the latest inflation,
$N_k \sim 60$ gives $|\epsilon_1| = {1 \over 2 N_k} \sim 0.0083$.
Then, we derive
\bea
   & & r_2 \sim 0.0029, \quad
      {M \over m_{pl}} \sim 2.6 \times 10^{-6}, \quad
      {H(t_k) \over M} \sim 4.5.
   \label{result}
\eea
The result shows the  negligible gravitational wave
contribution to CMBR anisotropies.
Also, we can show that $\dot \epsilon_1 = - 2 \epsilon_1^2 $
and $\ddot H \sim {\cal O} (\epsilon_1^3)$, thus
$\epsilon_4 = \epsilon_1$ and eq. (\ref{n-result-f}) becomes
$n_S - 1 \simeq 4 \epsilon_1 \sim - {2 / N_k}$, thus
\bea
   & & n_S - 1 \simeq - 0.033, \quad n_T \simeq 0.00.
   \label{n-result}
\eea

\section{Conclusions}
                                             \label{sec:Discussions}

We show that $R^2$ gravity gives nearly scale-invariant
Harrison-Zel'dovich spectra
considering the slow-roll parameters to the linear order,
i.e., ${\cal O} (\epsilon_i)$.
The recent measurements of the CMBR by Boomerang and
Maxima-1 in small angular scale \cite{Bernandis:2000}
together with {\it COBE}-DMR data \cite{Bennett:1996} provide the
scalar spectral index $n_S -1 \simeq 0.01^{+0.17}_{-0.14}$ at the $95\%$
confidence level \cite{Jaffe:2000}.
Thus, the inflation model considered in this work can be a
generation mechanism
for the large-scale structures.
Also, for a successful inflation the gravitational wave
contribution to the CMBR anisotropies should be negligible compared with
the one of the scalar type perturbation.
We expect that future observations of the spectral
indices and the contribution
of the gravitational wave to CMBR temperature/polarization anisotropies
will test the considered inflation model.
The original version of this proceeding was published in
\cite{HN:2001}.



\Acknowledgements

This work was supported by grant No. 2000-0-113-001-3 from the Basic 
Research Program of the Korea Science and Engineering Foundation.

\end{document}